# Laser induced ultrafast Gd 4f spin dynamics in $Co_{100-x}Gd_x$ alloys by means of time-resolved XMCD.

T. Ferté[1], M. Beens[2], G. Malinowski[3], K. Holldack[4], R. Abrudan[4], F. Radu[4], T. Kachel[4], M. Hehn[3], C. Boeglin[1], B. Koopmans[2] and N. Bergeard[1]

[1] Université de Strasbourg, CNRS, Institut de Physique et Chimie des Matériaux de Strasbourg, UMR 7504, F-67000 Strasbourg, France.
[2] Department of Applied Physics, Eindhoven University of Technology, P.O. Box 513, 5600 MB Eindhoven, The Netherlands.
[3] Institut Jean Lamour, Université Henri Poincaré, Nancy, France.
[4] Helmholtz-Zentrum Berlin für Materialien and Energy GmbH, Albert-Einstein-Str. 15, Berlin, Germany.

**Abstract:**

We have studied the laser induced ultrafast quenching of Gd 4f magnetic order in ferrimagnetic $Co_{100-x}Gd_x$ alloys to highlight the role of the Co 3d – Gd 5d inter-atomic exchange coupling. We have taken advantage of the ultrashort soft X-ray pulses deliver by the femtoslicing beamline at the BESSY II synchrotron radiation source at the Helmholtz-Zentrum Berlin to perform element- and time-resolved X-ray Magnetic Circular Dichroism spectroscopy. Our results show that the laser induced quenching of Gd 4f magnetic order occurs on very different time-scales for the $Co_{72}Gd_{28}$, the $Co_{77}Gd_{23}$ and the $Co_{79}Gd_{21}$ alloys. Most of the magnetic moment losses occur within the first picosecond (ps) while the electron distribution is strongly out of equilibrium. After the equilibration of the electrons and lattice temperatures (t > 1 ps), the magnetic losses occur on slower rates that depend on the alloy composition: increasing the Co composition speeds up the demagnetization of Gd 4f sublattice. The strength of the Co 3d – Gd 5d inter-atomic exchange coupling which depends on composition, determines the efficiency of the angular momentum flow from the Gd 4f spin towards the lattice. Our results are in qualitative agreements with the predictions of the microscopic three temperatures model for ferrimagnetic alloys.

## 1. Introduction:

The fundamental researches in spintronics aim at providing faster and energy efficient devices for the processing, the transfer and the storage of numerical data [1]. It is particularly crucial to address the needs of tomorrow's digital society while complying with the environmental challenges. The intensive experimental and theoretical works carried out in the field of *"Femtomagnetism"* over the last 25 years [2 - 4] have provided new paradigms to reduce both the time and the energy required for deterministic spin manipulation in magnetic materials by using infrared (IR) femtosecond (fs) laser pulses [5]. All-Optical Helicity Independent Switching (AO-HIS) induced by a single IR fs laser (or electron) pulse [6 - 8] and ultrafast spin switching induced by a single femtosecond spin-polarized hot-electron (SPHE) pulse via the Spin-Transfer Torque (STT) [9 - 11] are among the most promising solutions within this scope. Interestingly, the ferrimagnetic transition metal (TM) / Gd alloys and multilayers play a central role in both processes. Several studies have established some state diagrams that have revealed a dependence of their switching efficiencies notably on the alloy composition [10, 12]. However, these studies are mainly based on time-resolved Magneto-Optical Kerr Effect (TR-MOKE) which is only sensitive to the laser induced ultrafast TM 3d spin dynamics in these GdTM alloys. Previous experiments based on element- and time-resolved X-ray Magnetic Circular Dichroism (TR-XMCD) have reported on distinct Gd 4f and TM 3d spin dynamics induced by IR fs laser pulses in such Gd based alloys [13 - 18]. These distinct laser induced ultrafast spin dynamics (demagnetization amplitudes and characteristic demagnetization times) are believed to be the key parameters to achieve the ultrafast spin switching either by AO-HIS or ultrafast STT [9, 13]. It has recently been shown that the Gd 4f spin angular momentum is transferred to the lattice via the intra-atomic Gd 5d – Gd 4f and the inter-atomic Gd 5d – TM 3d exchange couplings [18] since there is no direct Gd 4f spin-lattice coupling [19 - 22]. Because of the localized nature of the Gd 4f electrons, the intra-atomic Gd 5d – Gd 4f exchange coupling is believed to be independent on alloy composition. Oppositely, it is believed that the inter-atomic Gd 5d – TM 3d exchange coupling depends on the composition and that its strength has an influence on the Gd 4f spin dynamics [23]. As a consequence, it is of paramount importance to extract experimentally the Gd 4f spin dynamics as a function of the inter-atomic Gd 5d – TM 3d exchange coupling (and thus on composition) to interpret the state diagrams and to reveal the mechanisms behind these single-shot spin switching [10, 12]. This knowledge is still fragmented because of the lack of systematic experimental investigations although theoretical models have attempted to fill this gap [23 - 25, 28].

In this preliminary work, we have investigated the laser induced ultrafast Gd 4f spin dynamics in $Co_{100-x}Gd_x$ alloys with $0.2 < x < 0.3$ by means of element- and time-resolved X-ray Magnetic Circular Dichroism (tr-XMCD) [13 - 16, 18]. We have chosen these alloys because they display a sharp variation of the Curie temperature ($T_{Curie}$) with the composition [26, 31]. Since $T_{Curie}$ is determined by the inter-atomic Gd 5d - Co 3d exchange coupling in these alloys [25], it allows studying the role of exchange coupling on the laser induced ultrafast Gd 4f spin dynamics in a narrow composition range. Our results show that the laser induced quenching of Gd 4f magnetic order occurs on very different time-scales for the $Co_{72}Gd_{28}$, the $Co_{77}Gd_{23}$ and the $Co_{79}Gd_{21}$ alloys. We evidenced two steps demagnetization for the $Co_{72}Gd_{28}$ and the $Co_{77}Gd_{23}$ alloys. Most of the magnetic moment losses occur within the first ps while the electron distribution is strongly out of equilibrium. After the equilibration of the electrons and lattice temperatures, the magnetic losses occur on much slower rates that depend on the alloy composition. We show that increasing the Co composition speed up the Gd 4f demagnetization rate which is qualitatively consistent with our calculations based on the microscopic three temperatures (m3T) model for localized 4f spins [27 - 29] and the scenario proposed by Eschenlohr et al. in $Tb_{100-x}Gd_x$ alloys [20].

## 2. Experimental methods, samples preparation and characterization:

The $Co_{100-x}Gd_x$ (20) alloy layers were grown by DC-magnetron sputtering on $[Ta(5)/Cu(20)/Ta(5)]_{x3}$ multilayers deposited on $Si_3N_4$ (200) membranes (unit in nm). The alloy layers were capped with Al(5) to prevent oxidization in ambient air [30]. We have selected alloys with x = 28, x = 23 and x = 21 that have Curie temperature ($T_{Curie}$) equal to 650 K, 765 K and 850 K respectively [26, 31]. We have characterized the dependence of their magnetic properties as a function of temperature by means of static X-ray Absorption Spectroscopy (XAS). These measurements have been performed by using the ALICE reflectometer installed on the PM3 beamline at the BESSY II synchrotron radiation source operated by the Helmholtz-Zentrum Berlin [32]. For these measurements, the magnetic field was applied along the X-rays propagation vector while the samples were tilted by 30° because of their in-plane uniaxial magnetic anisotropy [30, 33]. Hysteresis loops for the $Co_{72}Gd_{28}$ (figure 1a), the $Co_{77}Gd_{23}$ (figure 1b) and the $Co_{79}Gd_{21}$ (figure 1c) alloys were recorded by monitoring the transmission of circularly polarized X-rays tuned to the resonant Co $L_3$ edge as a function of the external magnetic field. The dependence of the coercive field on temperature for the three samples is shown in figure 1d. The X-ray Absorption Spectra (XAS) at the Co $L_3$ (figure 2a) and Gd $M_5$ (figure 2b) edges were acquired by monitoring the transmission of circularly polarized X-ray under magnetic fields of ± 1 kOe. The XMCD amplitude (in %) is defined as the difference divided by the mean value of the X-ray absorption at the resonance for two opposite field directions. The XMCD amplitude as a function of temperature at the Co $L_3$ and Gd $M_5$ edges for the $Co_{72}Gd_{28}$, the $Co_{77}Gd_{23}$ and the $Co_{79}Gd_{21}$ alloys are shown in figures 2a, 2b and 2c respectively.

The laser induced ultrafast Gd 4f spin dynamics was recorded by using element- and time-resolved X-ray Magnetic Circular Dichroism (tr-XMCD) at the femtoslicing beamline of the BESSY II synchrotron radiation source at the Helmholtz-Zentrum Berlin [34]. The transient XMCD signals have been measured by monitoring the transmission of circularly polarized X-ray pulses tuned to specific core level absorption edges as a function of a pump-probe delay for two opposite directions of the magnetic field. The photon energy was set to the Gd $M_5$ edges using a reflection zone plate monochromator on UE56/1-ZPM. A ~ 500 µm (FWHM) beam diameter for the pump laser on the sample was selected in order to ensure homogeneous pumping over the probed area of the sample (~200 µm). A magnetic field of ±5.5 kOe was applied along the propagation axis of both the IR laser and the X-rays beam during the experiment. Its strength is larger than the coercive fields of the $Co_{100-x}Gd_x$ alloys in the explored temperature range (figure 1d). As for the static XMCD measurements, the samples were tilted by 30° with respect to the X-ray propagation axis. The temperature of the cryostat was set to T = 80 K while the laser power was tuned to reach approximately ~ 45% demagnetization amplitude for the $Co_{72}Gd_{28}$ and the $Co_{77}Gd_{23}$ alloys (figure 3a and 3b). We have worked in a "moderate" excitation regime to avoid laser induced damages of the layers and to match the demagnetization amplitudes reported for pure Gd layer in the pioneer work of Wietstruk et al [19]. Unfortunately, the tr-XMCD measurements at T = 80 K were not possible for the $Co_{79}Gd_{21}$ alloy since the transient lattice temperature exceeded its temperature of magnetic compensation ($T_{comp}$ ~ 250 K) resulting in heat-assisted magnetization reversal (figure 4) [36]. The temperature of the cryostat was then set to T = 300 K (thus above $T_{comp}$) for this sample (figure 3c). It is worth noticing that $T^* = | T - T_{Curie} |$ is very similar for the measurements on the $Co_{72}Gd_{28}$ at T = 80 K ($T^* \sim 570$ K) and the $Co_{79}Gd_{21}$ alloys ($T^* = 550$ K) [35].

## 3. Experimental results and discussion:

By considering the characterization of the static magnetic properties of our samples, we observed a pronounce increase of the coercive field between T = 220 K and T = 270 K for the $Co_{79}Gd_{21}$ alloy (figure 1d). It is accompanied by a reversal of the hysteresis loop's sign when the temperature is increased from T = 220 K to T = 270 K (figure 1c) [37, 38]. This change of sign above 220 K, which is also visible in the XMCD signals at the Gd $M_5$ and Co $L_3$ edges (figure 2c), shows that the temperature of magnetic compensation is $T_{comp}$ ~ 250 K for the $Co_{79}Gd_{21}$ alloys. This value for $T_{comp}$ is consistent with the one tabulated for the same composition in literature [31]. For the $Co_{72}Gd_{28}$ and the $Co_{77}Gd_{23}$ alloys, we do not observe any reversal of the hysteresis loops nor any sign changes of the XMCD signals which means that $T_{comp}$, if it exists for these alloys, is above T = 300 K. Nevertheless, the coercive field slightly increases with the temperature for the $Co_{77}Gd_{23}$ alloy which indicates that $T_{comp}$ is slightly above T = 300 K. It is also consistent with the tabulated data in literature [31]. The static magnetic properties of the alloys attest that the nominal compositions are close (~ ±1%) to the actual ones. Interestingly, the XMCD amplitudes at the Gd $M_5$ edge are very similar for the $Co_{72}Gd_{28}$, the $Co_{77}Gd_{23}$ and $Co_{79}Gd_{21}$ alloys and they show almost identical temperature dependence (figure 2f). It shows that the average magnetic moment per Gd atoms ($\mu_{Gd}$ in $\mu_B$/atoms) is nearly identical in these alloys which is consistent with the localized nature of Gd 4f spins. In the following, we will consider that $\mu_{Gd}$(80 K) = 7 $\mu_B$/atoms for all samples which is consistent with mean field calculations [30].

In figure 3a, 3b and 3c, we show the transient XMCD as a function of the pump-probe delay for the $Co_{72}Gd_{28}$, the $Co_{77}Gd_{23}$ and the $Co_{79}Gd_{21}$ alloys respectively. The data are normalized by the XMCD signal at negative delays. For the $Co_{79}Gd_{21}$ alloy, we have further normalized the data by the $\mu_{Gd}$(300 K) / $\mu_{Gd}$(80 K) (~0.7) ratio to account for the averaged Gd magnetic moment per atom at T = 300 K. We observe a laser induced partial quenching of the Gd 4f magnetic order followed by a recovery but the time-scales vary with the alloy composition. The recovery of Gd 4f magnetic order occurs on a much longer time scale for the $Co_{72}Gd_{28}$ alloy (we extracted a characteristic recovery time $\tau_{Recov}$ ~ 160 ± 48 ps) than for the $Co_{79}Gd_{21}$ alloy ($\tau_{Recov}$ ~ 43 ± 8 ps). The characteristic recovery times reflect the rate at which the heat injected by the fs laser pulses is evacuated from the magnetic layers. The thermal heat conductivity (considering both electron and phonon transports) is strongly temperature dependent for ferrimagnetic alloys [39]. The measurements were carried out at T = 80 K for the $Co_{72}Gd_{28}$ alloy (figure 3a) and T = 300 K for the $Co_{79}Gd_{21}$ alloy (figure 3c). Therefore, we infer that the different base temperatures contribute to the different recovery times.

The maximum demagnetization amplitudes are $A_{28}$ = 0.45 ± 0.04, $A_{23}$ = 0.4 ± 0.04 and $A_{21}$ = 0.58 ± 0.03 while the delay at which these maximum demagnetization states are reached are $d_{28}$ = 12.5 ± 1 ps, $d_{23}$ = 7.5 ± 0.5 ps and $d_{21}$ = 3.5 ± 0.5 ps for the $Co_{72}Gd_{28}$, the $Co_{77}Gd_{23}$ and the $Co_{79}Gd_{21}$ alloys respectively (figure 3). It is worth noticing that the $Co_{72}Gd_{28}$ and the $Co_{77}Gd_{23}$ alloys display almost identical demagnetization amplitude and magnetic moment per Gd atoms. Therefore, they exhibit an almost identical loss of magnetic moment per Gd atoms ($\Delta \mu_{Gd}$ = 3.15 ± 0.3 $\mu_B$/at and 2.8 ± 0.3 $\mu_B$/at respectively). As a consequence, the demagnetization rate, which we define as $\Delta\mu_{Gd}$/d, is enhanced in the $Co_{77}Gd_{23}$ alloy as compared to that in the $Co_{72}Gd_{28}$ alloy (figure 5). Furthermore, in spite of larger demagnetization amplitude, the $Co_{79}Gd_{21}$ alloy also shows an almost identical loss of its magnetic moment (2.84 ± 0.2 $\mu_B$/at) as compared to the $Co_{72}Gd_{28}$ and the $Co_{77}Gd_{23}$ alloys but its demagnetization rate is further enhanced (figure 5). We have also displayed the transient normalized XMCD signal during the few picoseconds past the laser excitation in figure 3d, 3e and 3f. We observe that the demagnetization of the Gd 4f sublattice occurs on two subsequent time-scales for the $Co_{72}Gd_{28}$ and the $Co_{77}Gd_{23}$ alloys as previously reported for pure Gd layers [19, 27]. So far, such two-step demagnetization for the RE 4f sublattice in ferrimagnetic RE-TM alloys were experimentally observed in Fe(Co)Tb alloys [21, 40 - 43] by means of element-selective time-resolved MOKE [44] and indirectly in FeCoGd alloys [23, 24]. On contrary, several direct measurements by element- and time-resolved XMCD experiments have reported on "fast" single step demagnetization for RE 4f sublattices in $Co_{100-x}Dy_x$ and $Co_{100-x}Tb_x$ alloys [14, 16, 35, 45, 47]. It can be explained by sizable contributions of direct Dy 4f and Tb 4f spin-lattice coupling [21, 22, 48]. For sake of comparison, we have performed calculations based on the m3T model adapted to describe the spin dynamics in ferrimagnetic alloys [28]. We have injected the same input parameters in these calculations as those used in a previous study on alike $Co_{100-x}Gd_x$ layers [29]. The laser power was selected to match the experimental demagnetization amplitude while a sample composition x = 75 was chosen. The electronic and lattice temperatures as well as the transient magnetization are shown in figure 6. The simulated curves show a two steps demagnetization: a "fast" demagnetization step which occurs while the electron and the lattice temperatures are not thermalized (t < 1 ps) and a "slow" demagnetization step after the equilibration. In our experimental data (figure 3), we also observe that most of the Gd 4f magnetic moment losses occur within 1 ps for the $Co_{72}Gd_{28}$ and the $Co_{77}Gd_{23}$ alloys which is consistent with the characteristic electron-lattice equilibration times (figure 6). Since the Gd 4f spins are not directly excited by the fs laser pulses and in absence of Gd 4f direct spin-lattice coupling, this sub-picosecond quenching of Gd 4f spin order is necessarily driven by the photoexcited conduction band and by the intra-atomic Gd 5d – Gd 4f exchange coupling. After 1 ps past the laser excitation, the magnetization is still decreasing but at lower rates as compared to that in the excited states for both alloys. The dynamics is much slower for the $Co_{72}Gd_{28}$ alloy ($d_{28}$ ~ 12.5 ps) than for the $Co_{77}Gd_{23}$ alloy ($d_{23}$ ~ 7.5 ps). In the frame work of the m3T model, the characteristic demagnetization times scale with $\mu / (T_{curie})^2$ [27]. Since we have reported almost identical magnetic moments for the Gd atoms in the $Co_{72}Gd_{28}$ and the $Co_{77}Gd_{23}$ alloys (figure 2f), the different dynamics we observe can be qualitatively understood by considering the different $T_{Curie}$ and thus the different inter-atomic Gd 5d – Co 3d exchange coupling [25]. Wietstruk et al. have proposed that the second and "slow" demagnetization step in pure Gd layers is caused by the indirect Gd 4f spin-lattice coupling which is mediated by the Gd 5d – Gd 4f intra-atomic exchange coupling and the Gd 5d spin-lattice coupling [19]. In our $Co_{100-x}Gd_x$ alloys, the Gd 5d – Co 3d inter-atomic exchange coupling and the Co 3d spin-lattice coupling enhance the indirect Gd 4f spin-lattice coupling which speeds up the Gd 4f demagnetization in respect to pure Gd layer. Similar mechanism was proposed by Eschenlohr et al. in $Tb_{100-x}Gd_x$ alloys [20]. However, in their case, the indirect Gd 4f spin-lattice coupling was enhanced by the Gd 4f – Tb 4f RKKY coupling [49] and the direct Tb 4f spin-lattice coupling [20]. Here, we confirm that the Co spin-lattice coupling is an efficient channel for angular momentum transfer in ferrimagnetic alloys [18, 50] and this mechanism cannot be neglected even in case of "strong" direct RE 4f spin-lattice coupling [21].

For the $Co_{79}Gd_{21}$ alloy, we observed a single demagnetization step followed by a fast recovery (figure 3c, 3f). However, the composition is very similar to that of the $Co_{77}Gd_{23}$ alloy and we have used the very same laser power. As a consequence, we could have expected similar characteristic times for the equilibration of the electrons and lattice temperatures. Differences between the "fast" and "slow" time scales for the $Co_{79}Gd_{21}$ alloy are probably small and blurred by our experimental statistic. In addition, the faster demagnetization for the $Co_{79}Gd_{21}$ alloy ($d_{21}$ ~ 3.5 ps) than for the $Co_{77}Gd_{23}$ ($d_{23}$ ~ 7.5 ps) alloys is also consistent with its higher $T_{Curie}$ and enhanced Gd 4f spin-lattice coupling. Lopez et al. have reported on an almost complete demagnetization of the Gd 4f sublattice within 2 ps in a $Co_{80}Gd_{20}$ alloy [14]. The strong dependence of the characteristic demagnetization times on the composition is due to the sharp variation of the inter-atomic Co 3d – Gd 5d exchange coupling [25] which is attested by the sharp variation of $T_{Curie}$ with the composition [26]. Our experimental results show unambiguously the key role played by the inter-atomic Co 3d – Gd 5d exchange coupling in the ultrafast Gd 4f spin dynamics in $Co_{100-x}Gd_x$ alloys.

**4. Conclusions:**

We have shown that the laser induced ultrafast Gd 4f spin dynamics depends substantially on the composition in $Co_xGd_{100-x}$ alloys. We have observed that the quenching of the Gd 4f magnetic order occurs on 2 subsequent time scales as previously predicted by the m3T model [27] and experimentally observed in pure Gd layers [19, 20]. The "fast" demagnetization step occurs while the electrons in the conduction band and the phonons (i.e the lattice) are out of equilibrium. The "slow" demagnetization step occurs while the electrons and phonons have reached their thermal equilibrium. The characteristic demagnetization times of this "slow" demagnetization depend on the composition. In absence of direct Gd 4f spin-lattice coupling, our experimental results show unambiguously the key role played by the inter-atomic Co 3d – Gd 5d exchange couplings on the ultrafast Gd 4f spin dynamics in $Co_{100-x}Gd_x$ alloys as recently claimed by Zhang et al [22]. Further experiments are needed to determine whether the bottleneck for angular momentum transfer away from the Gd 4f sublattice in $Co_{100-x}Gd_x$ alloys is related by the intra-atomic exchange coupling or by the Co 3d spin-lattice coupling [49]. Our results also hint that for Co contents > 79%, the Gd 4f demagnetization occurs on similar time scales as that of Dy 4f (Tb 4f) in $Co_{100-x}Dy_x$ ($Co_{100-x}Tb_x$) alloys even without direct 4f spin-lattice coupling [21, 45, 46]. Quantitative analyses are needed to determine the transfer rates in $Co_{100-x}Gd_x$, $Co_{100-x}Dy_x$ and $Co_{100-x}Tb_x$ alloys to highlight the efficiency of direct RE 4f spin-lattice for angular momentum transfer. Finally, our measurements show that the recovery of Gd 4f magnetic order is strongly affected by the lattice temperature and the heat conductivity [39]. As a conclusion, our experiments demonstrate that the Gd 4f spin dynamics (demagnetization and recovery) can be finely regulated in $Co_{100-x}Gd_x$ alloys by tuning the temperature and the composition. More systematic element- and time-resolved experiments are definitely needed in order to correlate the Gd 4f spin dynamics with the state diagrams of AO-HIS [12] and ultrafast STT [9].

**Acknowledgments:**

We are indebted for the scientific and technical support given by N. Pontius, Ch. Schüßler-Langeheine and R. Mitzner at the slicing facility at the BESSY II storage ring. The authors are grateful for financial support received from the following agencies: the French "Agence National de la Recherche" via Project No. ANR-11-LABX-0058_NIE and Project EQUIPEX UNION No. ANR-10-EQPX-52 and the EU Contract Integrated Infrastructure Initiative I3 in FP6 Project No. R II 3CT-2004-506008**.** This work was supported partly by the French PIA project "Lorraine Université d'Excellence", reference ANR-15-IDEX-04-LUE, by the Project Plus cofounder by the "FEDER-FSE Lorraine et Massif Vosges 2014-2020", a European Union Program and by the OVNI project from Region Grand-Est and by the MATELAS project institut Carnot ICEEL.

**Figures:**

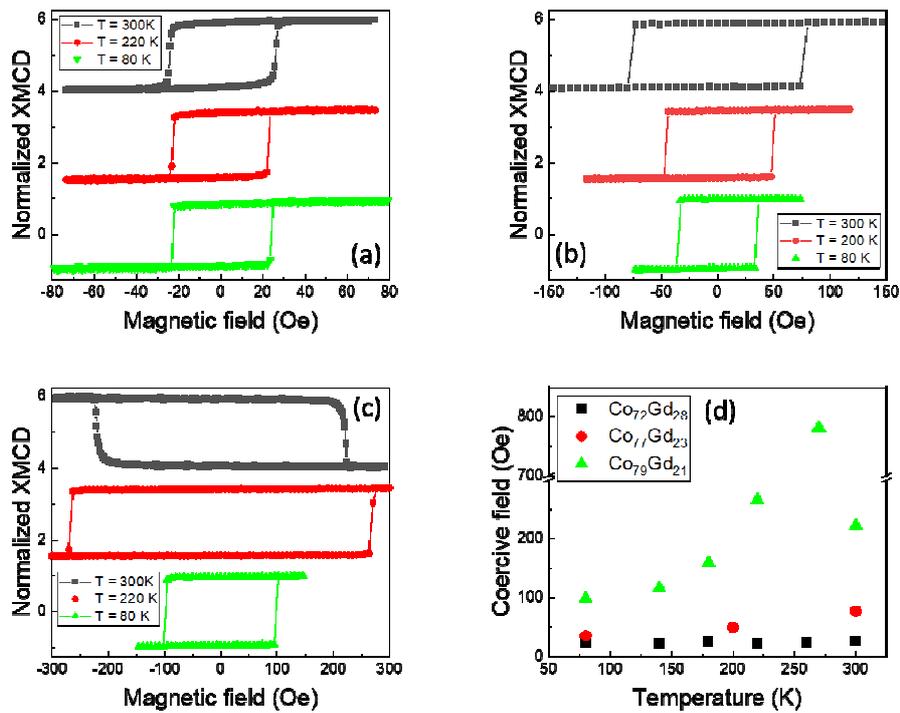

**Fig 1**: Hysteresis loops at selected temperatures recorded by monitoring the transmission of circularly polarized X-rays, tuned to match the resonance at the Co L$_3$ edge, as a function of the external magnetic field for the Co$_{72}$Gd$_{28}$ (a), the Co$_{77}$Gd$_{23}$ (b) and the Co$_{79}$Gd$_{21}$ (c) alloys. (d) Coercive field as a function of temperature (mind the break on ordinate axis).

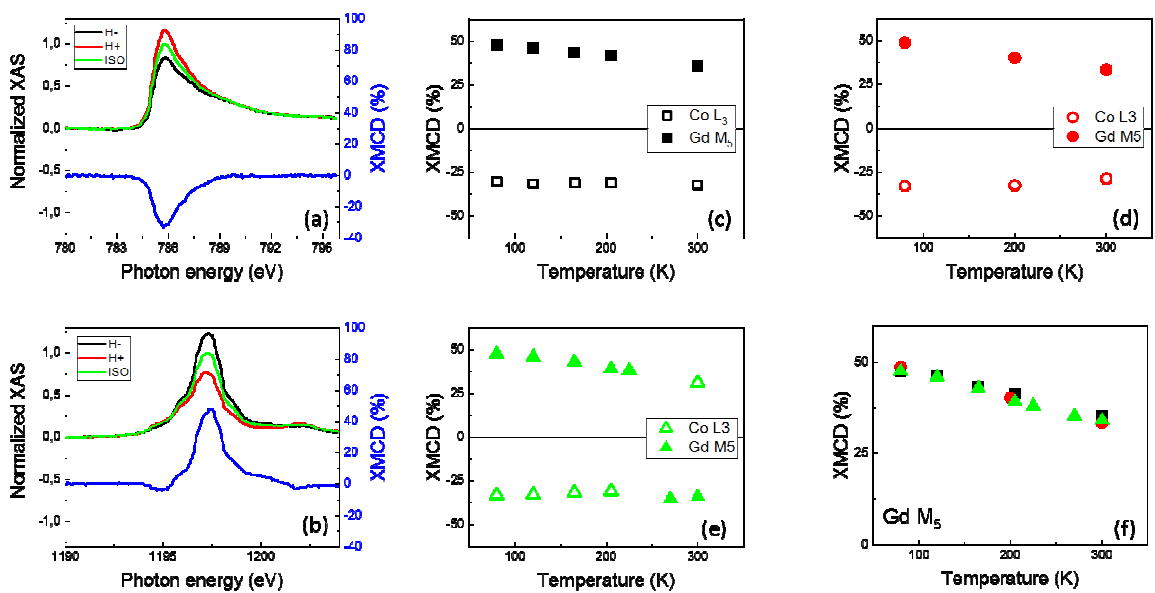

**Fig 2:** XAS and XMCD spectra recorded by monitoring the transmission of circularly polarized X-rays at the Co L$_3$ (a) and Gd M$_5$ (b) edges for the Co$_{72}$Gd$_{28}$ alloy at T = 80K. Maximum XMCD amplitude at the Co L$_3$ (open symbols) and Gd M$_5$

(filled symbols) edges as a function of temperature for the $Co_{72}Gd_{28}$ (c), the $Co_{77}Gd_{23}$ (d) and the $Co_{79}Gd_{21}$ (e) alloys. (f) Maximum XMCD amplitude at the Gd $M_5$ edge as a function of temperature for all alloys. The symbol sizes reflect the experimental uncertainty.

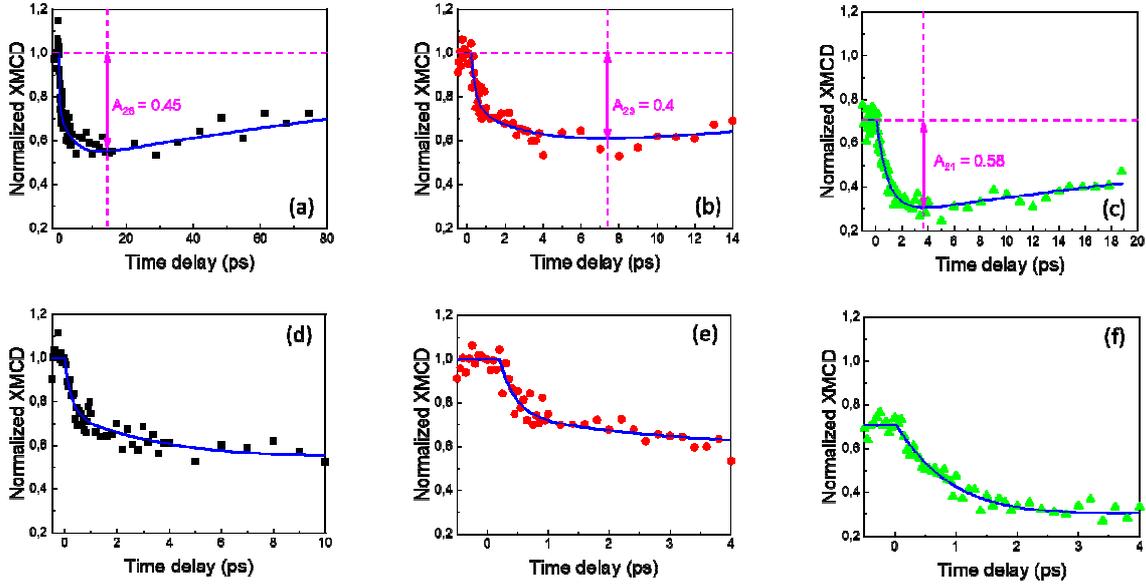

**Fig 3:** Transient XMCD at the Gd $M_5$ edges for the $Co_{72}Gd_{28}$ (a), the $Co_{77}Gd_{23}$ (b) and the $Co_{79}Gd_{21}$ (c) alloys as a function of the pump-probe delay. The measurements were carried out at T = 80 K for the $Co_{72}Gd_{28}$ and the $Co_{77}Gd_{23}$ alloys (figure 3a and 3b) and at T = 300 K for the $Co_{79}Gd_{21}$ alloy (figure 3c). The curves were normalized by the average XMCD amplitude at negative delays and at T = 80K (see text for details). The solid blue lines are exponential fits and serve as guides to the eyes. The maximum demagnetization amplitude ($A_x$) is labelled by the vertical arrows. (d, e and f) Same as (a, b and c) but on a narrower delay range.

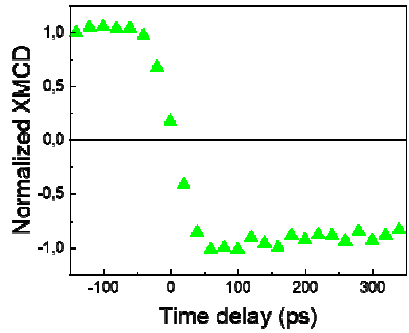

**Fig 4:** Transient normalized XMCD at the Gd $M_5$ edge measured in the hybrid mode (time resolution ~60 ps) for the $Co_{79}Gd_{21}$ alloy at T = 80K and laser power P = 70mW.

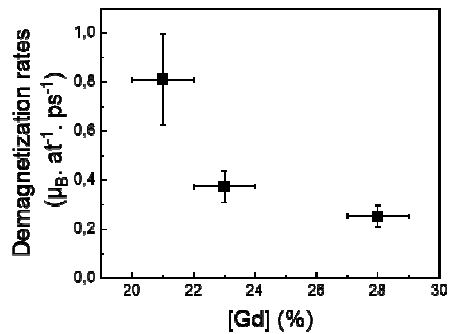

**Fig 5:** Demagnetization rates as a function of the Gd composition.

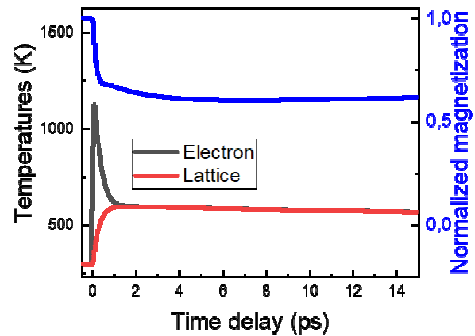

**Fig 6:** Normalized magnetization (blue) as well as the electronic (black) and the lattice (red) temperatures as a function of the time delay as calculated by the microscopic 3 temperatures model adapted for ferrimagnetic alloys [28, 29].